# A plasmonic walker

Chao Zhou, Xiaoyang Duan, and Na Liu[*]

*Max Planck Institute for Intelligent Systems, Heisenbergstrasse 3, D-70569 Stuttgart, Germany*
[*]*To whom the correspondence should be addressed. E-mail: laura.liu@is.mpg.de*

**In nano-optics, a formidable challenge remains in precise transport of a single optical nano-object along a programmed and routed path toward a pre-defined destination. In living cells, molecular motors such as kinesin and dynein can walk directionally along microtubules to ferry cargoes. Such biological machines have been the inspiration for the development of artificial molecular walkers, which are made of DNA strands or small molecules. Here, we demonstrate the first plasmonic walker, which can execute directional, progressive, and reverse nanoscale walking on two or three-dimensional DNA origami. The plasmonic walker comprises an anisotropic gold nanorod as its 'body' and discrete DNA strands as its 'feet'. Specifically, our plasmonic walker carries optical information and can *in situ* optically report its own walking directions and consecutive steps at nanometer accuracy, through dynamic coupling to a plasmonic stator immobilized along its walking track. The dynamic process can be optically read out at visible frequencies in real time. Our concept will enable a variety of smart nanophotonic platforms for studying dynamic light-matter interaction, which requires controlled motion at the nanoscale well below the optical diffraction limit.**

A plasmonic walker is an active plasmonic device that mimics the directional movement of naturally occurring molecular motors. Different from other artificial analogues such as DNA walkers, a plasmonic walker can be endowed with novel optical functionalities. A plasmonic walker works not only as a walking element to carry out mechanical motion but also as an optical reporter, which can deliver its own translocation information through optical spectroscopy in real time. This may leverage the scope of synthetic molecular machinery.[1-13] Also, it circumvents the complexity of conventional walker characterization techniques as well as allows for noninvasive and stable characterizations over long periods of time.

The key idea is to create a plasmonically coupled system, in which a walker and a stator, *i.e.*, an immobilized plasmonic element (or elements) on a walking track constitute a conformationally sensitive geometry.[14] When the walker carries out stepwise movements, it triggers a series of conformational changes of the system as well as activates subsequent near-field interaction changes with the stator, thus giving rise to immediate spectral response changes that can be read out optically. As a result, locomotion on the order of several nanometers, which is far below the optical resolution limit, can be optically discriminated in real time.

**Results**

As shown in Figure 1, a walker, gold nanorod (AuNR) in yellow and a stator (AuNR in red) are organized in a chiral geometry. Specifically, the walker and the stator are placed on two opposite surfaces of a two-dimensional (2D) rectangular DNA origami platform, forming a 90° cross configuration. A chiral geometry is chosen in that circular dichroism (CD), that is, differential absorption of left-and right-handed circularly polarized light, of three-dimensional (3D) chiral structures are markedly sensitive on their conformational changes.[15-18] The stator is immobilized on one surface through the capture strands on the origami, whereas on the other surface the walker can execute stepwise movements by programmably attaching (detaching) its feet on (from) the track through hybridization (de-hybridization) with the footholds (coded A-F in Fig. 1). In particular, double-layer DNA origami is utilized to achieve a rigid and robust track. The DNA origami (58 nm×42 nm×7 nm) was prepared by folding a long single-stranded DNA scaffold with staple strands and specific capture strands, following a self-assembly process (see Supplementary Note 1)[19,20].

In contrast to previous tiny DNA walkers,[1-11] our walker comprises an anisotropic AuNR, which is as large as 35 nm×10 nm. On one hand, a large metal nanoparticle is essential for plasmonic probing, as it yields distinct and pronounced optical response. On the other hand, the anisotropic nature of the AuNR also brings about substantial challenges to implement directional and progressive walking.

To impose directional walking, the feet of the walker and the footholds on the track are specifically designed. The walker AuNR is fully covered with identical foot strands, which contain a 9-nt segment for hybridization and four thymine bases as spacer. Along the track, six parallel rows of footholds A-F, are utilized to establish five walking stations I-V, which are evenly separated by 7 nm. This also defines the step size of the walker. At each station, the walker's feet step on two rows of the footholds to accommodate its transverse dimension as well as to ensure stable binding. In each row, five binding sites with identical footholds are extended from the origami. Each foothold consists of two parts: a binding segment (9-nt, black) for hybridization with a foot strand of the walker as well as a toehold segment (8-nt, colored), which is differently sequenced in different foothold rows for achieving programmable reactions.

Fig. 2a schematically describes the walking principle. Initially, the walker resides at station I (start site), stepping on rows A and B through DNA hybridization. This was implemented with the assembly of the stator on the origami during the same annealing process (see Supplementary Note 2). Foothold rows C, D, E, and F are deactivated by respective blocking strands. At station I, the walker and the stator form a left-handed configuration. Due to close proximity, the two AuNRs can be strongly coupled. This generates a theme of handedness when interacting with left and right circularly polarized light, giving rise to CD.[17,21-23] To correlate discrete walking steps and their associated optical response, CD spectra at different walking stations were measured using a Jasco-815 CD spectrometer. All the measurements were carried out at room temperature and pH 8.0. The CD spectrum at station I is presented by a green curve in Fig. 2b, showing a characteristic peak-to-dip line shape centered around 740 nm. The measured CD intensity is as large as 200 mdeg at a sample concentration of approximately 0.67 nM. Such strong and distinct spectral response enables highly sensitive spectroscopy, which is the basis to optically monitoring structural dynamics.

The stepwise walking is powered by DNA hybridization and activated upon addition of respective blocking and removal strands. The blocking and removal strands for footholds A-F are labeled as *a-f* and $\bar{a}$-$\bar{f}$, respectively. Each blocking strand consists of three parts: the upper (11-nt, colored), the middle (6-nt, black), and the bottom segments (8-nt, colored). The removal strands are fully complementary to their corresponding blocking strands (see Supplementary Notes 3 and 4). First, blocking strands *a* and removal strands $\bar{c}$ are added simultaneously. Dissociation of the walker's feet from row A is initiated by blocking strands *a* through a strand-displacement reaction mediated by the toehold segments on footholds A. Row A is then site-blocked. This eliminates the back stepping of the walker, thus imposing directionality. It is crucial to underline that blocking strand *a* is specifically designed so that the toehold segment of foothold A is fully hybridized with the bottom segment of blocking strand *a*, whereas the binding segment of foothold A is only partially hybridized with the middle segment of blocking strand *a*, leaving 3-nt of the foothold unpaired. This ensures the specificity of the blocking strands for different foothold rows and more importantly avoids undesirable hybridization between the removal strands and the walker's foot strands (see Supplementary Note 3).

Meanwhile, blocking strands *c* are dissociated from row C by removal strands $\bar{c}$ through branch migration, triggered by the upper segments of blocking strands *c* as toeholds. Row C is then activated. The walker's feet search for an accessible neighboring site and subsequently bind to footholds C (see Supplementary Note 4). As a result, the walker executes one step forward and reaches station II by stepping on rows B and C as shown in Fig. 2a. It is worth mentioning that during the process of detaching from row A and attaching to row C, one set of the walker's feet stay bound to row B, preventing the walker from being off the track. In a more descriptive picture, the walker imposes directional walking by alternatively advancing its feet in a 'rolling' fashion. At station II, the walker and the stator form a less asymmetric configuration compared to the case at station I. This leads to an immediate CD decrease as presented by the blue curve in Fig. 2b, indicating forward motion of the walker.

To impose progressive walking, blocking strands $b$ and removal strands $\bar{d}$ are added subsequently. Following a similar aforementioned principle, the walker executes one further step forward, reaching station III by stepping on rows C and D. At this station, the CD spectrum in principle should exhibit no spectral features in that the walker should nominally stand along the central axis of the stator, forming an achiral configuration. However, as presented by the brown curve in Fig. 2a, a slight right-handed preference is visible in the CD spectrum. This is possibly due to the assembly imperfection in the experiment as a minute deviation of the walker from the central axis of the stator can lead to immediate CD signals resulting from the high sensitivity of CD spectroscopy. Sequential addition of corresponding blocking and removal strands enables progressive walking further toward stations IV and V. The walker enters the right-handed configuration region. As shown in Fig. 2b, when the walker strides from station III to IV and subsequently to V, the CD response strengthens successively. At station V, the CD response reaches approximately -200 mdeg, exhibiting a dip-to-peak line shape, which is nearly a mirror image of the CD spectrum at station I. This importantly indicates that in the solution the walkers that were directed to walk from station I have nearly all successfully reached station V, demonstrating the high fidelity of the walking process. In short, the individual steps of the walker which are well below the optical diffraction limit can be optically discriminated in real time.

For comparison, theoretical calculations[24] of the CD spectra were carried out (for details, see Supplementary Note 5) and are presented in Fig. 2c. In the calculations, the right-handed preference at station III was not included. Overall, the experimental spectra agree well with the theoretical results. In addition, to assess the assembled nanostructures, transmission electron microscopy (TEM) was performed. TEM images of the DNA origami templates and exemplary structures at station I are shown in Figs. 3a and 3b, where the rectangular origami and the formation of AuNR dimers are clearly visible. Enlarged TEM images of the AuNRs assembled on the origami from different perspectives are presented in Fig. 3c. TEM images of the structures at other stations can be found in Supplementary Note 4.

To *in situ* monitor the dynamic walking process, CD spectra of the sample were recorded using a time-scan function of the CD spectrometer at a fixed wavelength of 685 nm. As shown in Fig. 4, the CD intensity displays a successive decrease when the walker executes discrete steps from station I to station V. In average, the transition between different steps takes approximately 25 min to complete. Previous DNA walkers based on 'burnt bridge' render impossible reverse walking along the same track.[4,8,11] To demonstrate the switchable directionality of our walker, reverse walking is carried out after the walker reaches station V. Upon addition of blocking strands $f$ and removal strands $\bar{d}$, the walker changes its walking direction and executes one step back toward station VI, stepping on rows D and E. This gives rise to an instant CD intensity increase as shown in Fig. 4. When the walker executes one more step backward, the CD intensity shows a further increase to the level at station III. Subsequently, the walker makes a new turn at station III and undergoes another reverse walking toward station IV. As shown Fig. 4, the CD intensity changes approximately back to the level at station IV. Overall, the walker has successfully carried out directed movements along the track, following a regulated route of I-II-III-IV-V-IV-III-IV.

To demonstrate the capability to perform more complex behavior, stepwise walking of the plasmonic walker on a 3D origami platform is examined. Fig. 5a shows the schematic of the walker system, in which triangle-prism origami is utilized as the walking track. Its length is 35 nm. The three side-lengths of the triangular cross-section are approximately 29 nm, 26 nm, and 38 nm. The stator (AuNR in red) is immobilized on one side-surface of the triangle prism (see Fig. 5a). Seven parallel rows of footholds are extended from the other two side-surfaces to establish six walking stations I-VI. At each station, the feet of the walker (AuNR in yellow) step on two rows of the footholds. Detailed design information can be found in Supplementary Note S6. The stepwise walking starts from station I, where the walker and the stator form a right-handed configuration. Following the same walking strategy described in Fig. 2, the walker first carries out two discrete steps along the track, reaching stations II and III, respectively. Subsequently, the walker approaches the vertex of the triangle prism at station IV. It then makes a turn, entering the left-handed configuration region. By executing two more discrete steps, the walker eventually reaches station VI. Fig. 5b shows the experimental CD spectra at different walking stations. It is evident that the individual steps along the track on the 3D origami can be correlated with distinct CD spectral changes. Representative TEM images of the AuNRs on the 3D origami at different stations are also presented in Fig. 5b. The corresponding calculated CD spectra can be found in Supplementary note S6. The experimental and theoretical results show an overall good agreement.

The powerful combination between the precise control of nanoscale motion enabled by DNA nanotechnology and the rich spectral information offered by plasmonics[25-27] suggests a new generation of artificial synthetic machines, which can *in situ* report their own structural dynamics using a noninvasive, stable, and all-optical approach. This will render profound significance in dual disciplines. First, the realization of advanced plasmonic walkers that can stride along multidirectional footpaths and perform different taskson 2D or 3D prescriptive landscapes can be promptly envisioned.[8,28,29] Second, our walker concept will expand the functional scope of DNA-based devices as well as enrich the category of the state-of-the-art characterization methods for practical applications. Intriguing light-matter interaction studies, for example, distance-dependent interaction between single emitters and plasmonic nanoparticles will no longer be restricted to a static picture.[30-32] The plasmonic nanoparticle can be transformed to a walker by proper functionalization and prods the emitter with fully coordinated motion at the nanoscale accuracy. Finally, our walker concept also outlines an exciting prospect of generating programmable large-scale nanocircuits that incorporate biochemical, electrical, and optical components for active transport and information processing.

## Acknowledgements

We thank A. Jeltsch and R. Jurkowska for assistance with CD spectroscopy. We thank M. Kelsch and K. Hahn for assistance with TEM microscopy. We thank H. Ries for DNA scaffold preparation. We acknowledge S. Hein for material visualizations. TEM data was collected at the Stuttgart Center for Electron Microscopy (StEM). N.L. was supported by the Sofja Kovalevskaja Award from the Alexander von Humboldt-Foundation, a Marie Curie CIG Fellowship, the European Research Council (ERC *Dynamic Nano*), and the Grassroots Proposal M10330 from the Max Planck Institute for Intelligent Systems.

## Author contributions

C.Z. and N.L. conceived the concept. C.Z. designed the DNA origami nanostructures as well as walking principle and performed the experiments. X.D. carried out the theoretical calculations. C.Z. and N.L. wrote the manuscript. All authors discussed the results, analyzed the data, and commented on the manuscript.

## Additional information

Supplementary information is available in the online version of the paper. Reprints and permissions information is available online at www.nature.com/reprints. Correspondence and requests for materials should be addressed to N.L.

**Competing financial interests**

The authors declare no competing financial interests.

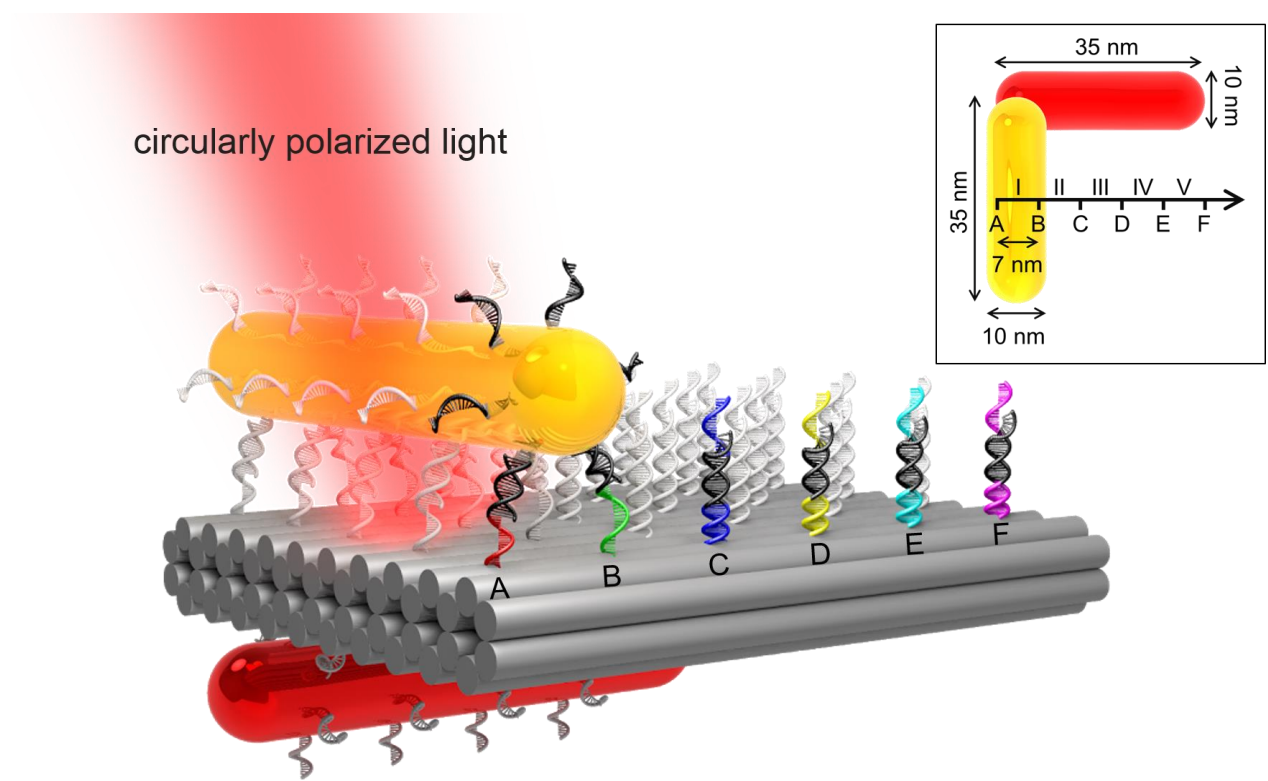

**Figure 1 | Schematic of the plasmonic walker.** Two gold nanorods (AuNRs) are assembled perpendicularly to one another on a double-layer DNA origami template, forming a left-handed configuration at station I. The yellow AuNR on the top surface represents the "walker" and the red AuNR on the bottom surface represents the "stator". The walking track comprises six rows of footholds (A-F) extended from the origami surface to define five walking stations (I-V). The distance between the neighboring stations is 7 nm, which also corresponds to the step size. In each row, there are five binding sites with identical footholds. Only the footholds in the front line are colored to highlight the different strand segments. The walker AuNR is fully functionalized with foot strands. To enable robust binding, the walker steps on two neighboring footholds at each station. The red beam indicates the incident circularly polarized light.

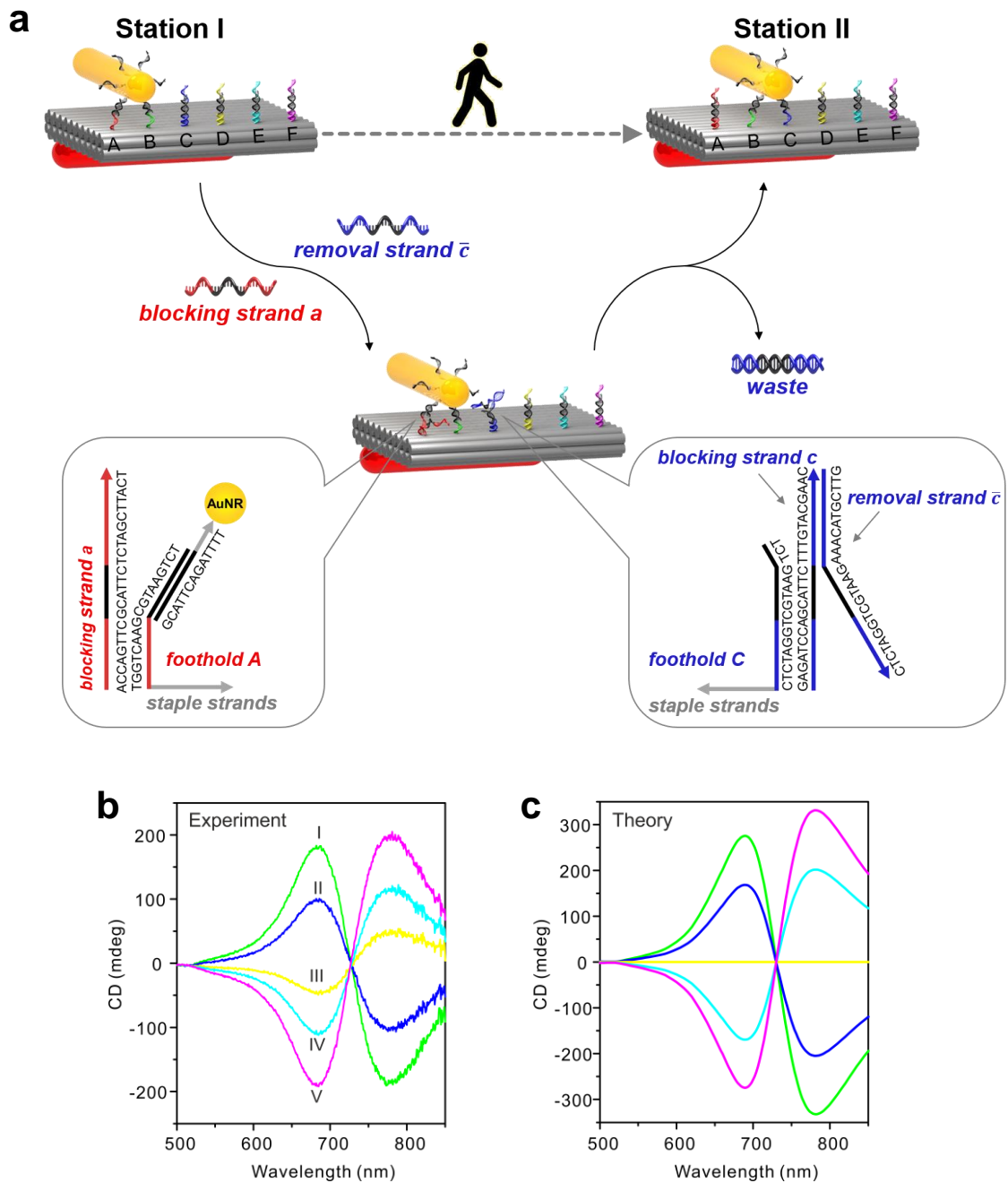

**Figure 2 | Walking mechanism, measured and simulated CD spectra at each station. a**, walking mechanism. Initially, the walker resides at station I. Upon addition of blocking strands *a* and removal strands $\bar{c}$, two toehold-mediated strand-displacement reactions occur simultaneously. Blocking strands *a* trigger the dissociation of the walker's feet from footholds A. Row A is then site blocked. Meanwhile, removal strands $\bar{c}$ release blocking strands *c* from footholds C. Row C is therefore site activated to bind the feet of the walker. Subsequently, the walker carries out one step forward, reaching station II. For simplicity, only the front line of the associated strands is shown. **b**, Measured CD spectra at each station. **c**, Simulated CD spectra at each station. The right-handed preference at station III was not included in the calculation.

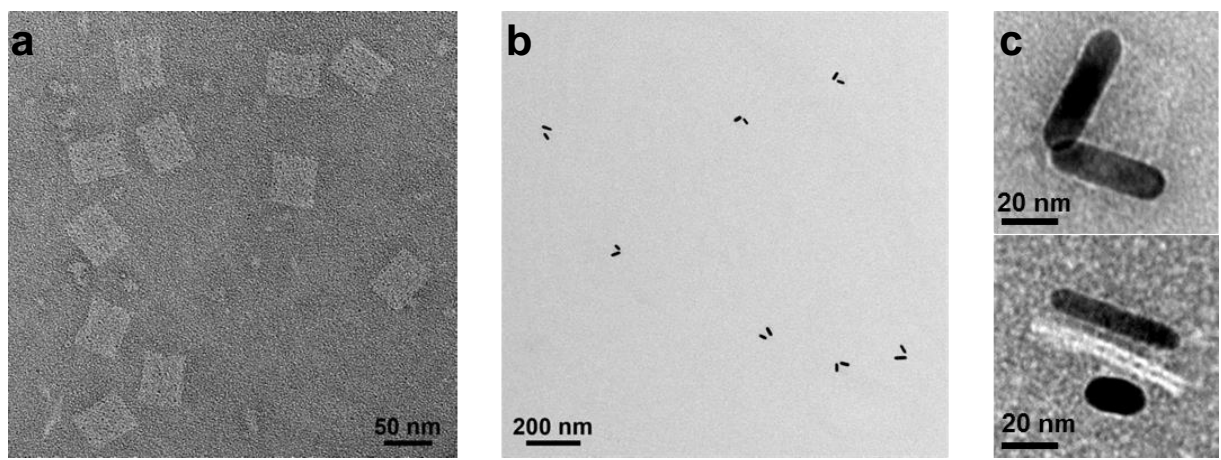

**Figure 3 | TEM images of the DNA origami templates and the plasmonic walker structures. a**, TEM image of the double-layer DNA origami (58 nm×42 nm×7 nm) after negative staining; **b**, Exemplary TEM image of the plasmonic walker structures at station I. In the individual structures, two AuNRs (35 nm×10 nm) are assembled on one origami template. The plasmonic structures display certain deformation due to the drying process on the TEM grids. **c**, Enlarged-views of the plasmonic walker structures from different perspectives.

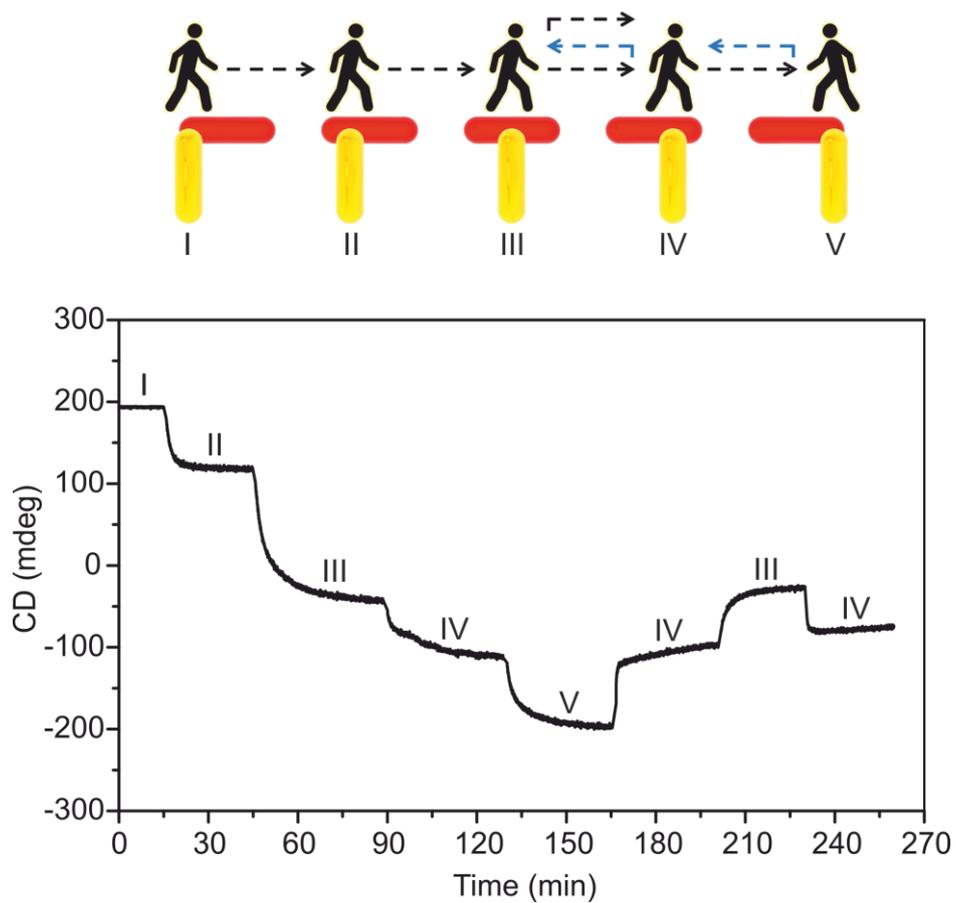

**Figure 4 | Directional and progressive walking of the plasmonic walker detected by in situ CD spectroscopy.** The CD intensity was monitored at a fixed wavelength of 685 nm, while the walker performs stepwise walking, following a route I-II-III-IV-V-IV-III-IV.

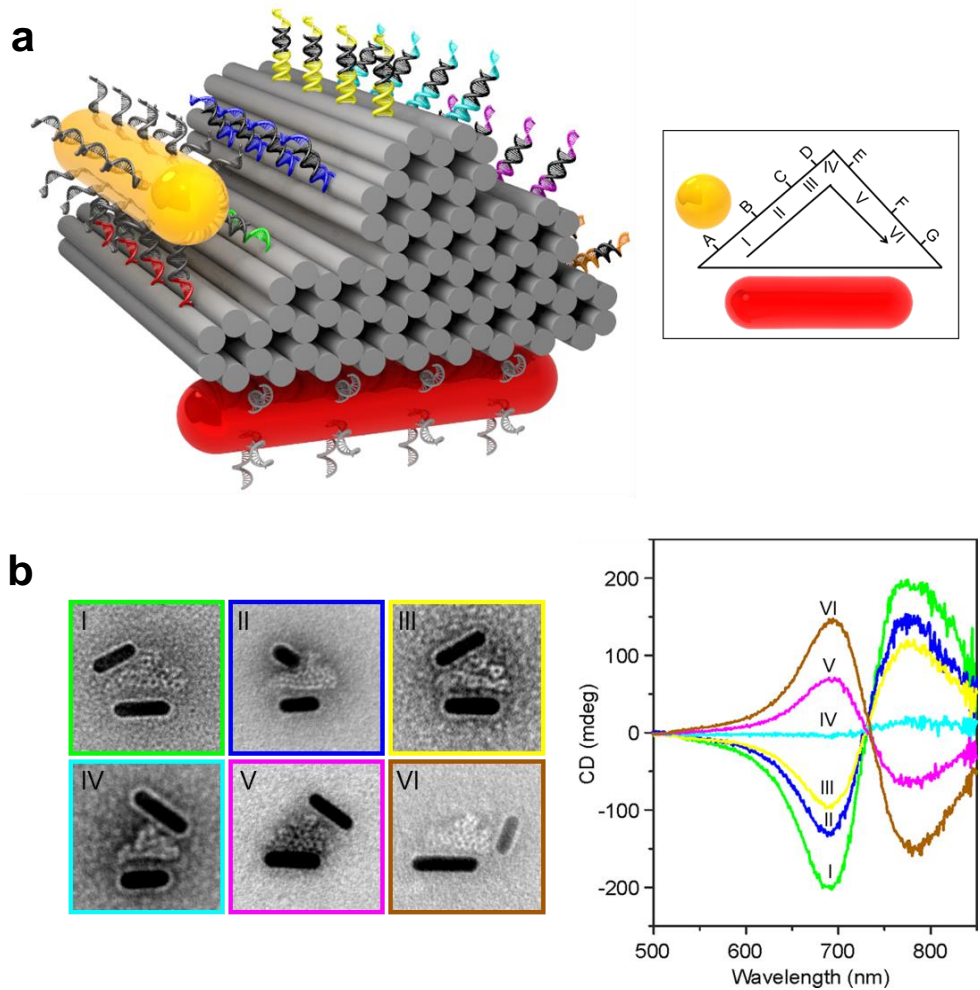

**Figure 5 | Stepwise walking on 3D DNA origami. a**, Schematic of the plasmonic walker on a 3D triangle-prism DNA origami platform. The walking track comprises seven rows of footholds (A-G) extended from the origami surface to define six walking stations (I-VI). The walking process starts at station I, where the walker and the stator form a right-handed configuration. The distances between the neighboring stations are slightly different owing to the irregular side-surfaces of the 3D origami. The successive step sizes are approximately 7 nm, 7 nm, 12 nm, 12 nm, and 11 nm. **b**, Measured CD spectra and corresponding TEM images of the plasmonic walker structures at different stations. The frame size of each TEM image is 80 nm. The plasmonic structures display certain deformation due to the drying process on the TEM grids.